\documentstyle[aps,pre,epsfig]{revtex}

\newcommand{\be}{\begin{equation}}
\newcommand{\ee}{\end{equation}}
\newcommand{\ba}{\begin{eqnarray}}
\newcommand{\ea}{\end{eqnarray}}

\title{Ring-driven shear thickening in wormlike micelles?}
 
\author{M. E. Cates$^1$ and S. J. Candau$^2$}
\address{
1. Department of Physics and Astronomy, University of Edinburgh, Kings Buildings, Mayfield Road, Edinburgh EH9 3JZ, Scotland, G.B.\\
2. Laboratoire de Dynamiques des Fluides Complexes, UMR7506,
Universit\'e Louis Pasteur, 4 rue Blaise Pascal, F-67070 Strasbourg, France\\
}

\begin{document}

\maketitle

\begin{abstract}
The shear-thickening behaviour of wormlike micelles, at concentrations just below ``overlap'', remains unexplained. In some cases, it has recently been confirmed that very slow relaxations must be present {\em even in the quiescent state} -- for example, the shear thickening properties can depend on earlier thermal cycling. We present a speculative scenario based on the presence, just below the (apparent) overlap threshold, of large rings whose linking and delinking kinetics control the shear thickening process. Equilibration between rings and open chains in turn controls the slow relaxations.

\end{abstract}

\section{Introduction}
Many surfactants in solution aggregate into micelles that are locally rodlike or wormlike in character. Well above an ``overlap threshold'' at volume fraction $\phi=\phi^*$, these are highly entangled and viscoelastic. The main features of are then explained by the reptation-reaction model, which couples the diffusive motion of entangled wormlike chains with the reaction kinetics (reversible scission, end-interchange or bond-interchange) of their micellar breakdown and reformation \cite{cc}. The model predicts extreme shear-thinning via a shear-banding mechanism \cite{spenley} now well confirmed experimentally \cite{thinex}, as is (in slightly different systems) a closely related scenario of shear-induced separation to a nematic phase \cite{nemband}. 

Completely altered physics is often observed just below the overlap threshold. Indeed, there are many systems, primarily ionic surfactants at low levels of added salt \cite{saltfree} (but also including at least one nonionic system \cite{matthys} and several more exotic `gemini' surfactants \cite{gemini}), in which dramatic shear thickening is observed in this region, as first studied by Hoffmann and coworkers \cite{hoffmann}. In many such cases, there is a rather abrupt (factor ten or more) jump in steady shear viscosity around a well defined threshold of shear rate ($\dot\gamma_c$).

Previous attempts to explain this, based on flow-induced aggregation of small rods or their alignment-induced end-to-end fusion \cite{previous,tb}, are somewhat unconvincing. For example, it is hard to explain why there is a gradual suppression of the shear thickening effect by adding salt \cite{suppression} (which if anything would lower the barriers to aggregation or fusion). It is also problematic that, if the only particles present for $\phi<\phi^*$ are micelles small enough not to overlap, then the longest relevant relaxation time in the quiescent system is their rotational diffusion time (Zimm time) $\tau_{z}$. This is typically micro- or milliseconds, whereas $\dot \gamma_c$ is normally of order $10-100$s$^{-1}$. In general, one would not expect to see a strong nonlinear effect of shearing until $\dot \gamma \tau\simeq 1$, where $\tau$ is the longest relaxation time of the quiescent system. 

This is true whether or not a new `phase' is created by shearing, unless such a phase (e.g. a nematic) is almost stable even without flow; near $\phi^* \simeq 0.01$ this is unlikely. Nucleation of a new phase would, however, help explain observations of a long latency period before reaching steady state and a long recovery after switchoff of shear \cite{hoffmann,latency,pine,Berret}. While in some cases the shear-induced gel does form a separate phase with nontrivial nucleation \cite{pine}, it may not always \cite{hoffmann}; and in any case, one still needs a microscopic explanation for the new phase. We explore below the possibility that slow relaxations at the structural level, arising from micellar rings, can explain the observed formation of a gel. In cases where phase separation also occurs, the additional phase kinetics could allow very complex behavior \cite{latency,pine,Berret}. 

Micellar rings are expected theoretically \cite{wheeler}: their apparent {\em absence} in most experiments has been mysterious \cite{catesring}. In the limit of a high end-cap energy $E/2$ for the micelles, the physics near $\phi^*$ is, according to theory, completely dominated by rings. When $E$ is infinite, there is at $\phi_c$ ($\simeq \phi^*$) a continuous ``polymerization transition'' where a power-law cascade of marginally overlapped rings emerges. (The size of the largest rings gives a correlation length $\xi$ which diverges smoothly at $\phi_c$.) For $\phi > \phi_c$ there is a condensate consisting of an infinitely long chain, alongside a dwindling ring population. Allowing for a finite $E$ one has, instead of the condensate, an entangled solution of long open chains; the divergence in $\xi$ is cut off and the transition is smoothed. (The distinction between a critical point $\phi_c$ and a smooth crossover to overlap at $\phi \simeq \phi^*$ is then lost.) For a low enough $E$, open chains are present well below $\phi^*$ and a ring-free scenario should suffice \cite{catesring}; this also applies far enough into the entangled region where linear chains always dominate. However many systems do have rather large $E$ values ($20-30kT$), and evidence for rings near $\phi^*$ has sometimes been found \cite{cryo}. 

\section{Experimental Features}
Let us recap what needs to be explained. In the following discussion, numerical parameters are taken from Ref.\cite{Berret};  we also refer to some new experiments which will be published separately elsewhere \cite{newexpts}.

The shear thickening transition is not discontinuous, but typically involves a factor ten increase in viscosity around a critical shear rate $\dot\gamma_c \simeq 50$s$^{-1}$. Below this threshold, the steady state viscosity is barely perturbed from its zero-shear value $\eta_0$. Above it, there is a latency time $\tau_{lat}$ between the onset of steady shear and the increase in stress associated with shear thickening. After this time, the sample is strongly and (in many cases) uniformly birefringent, with a near-zero extinction angle (and, when measured, a large first normal stress difference) implying the presence of a condensate of extremely aligned material. 
The latency time $\tau_{lat}$ gets larger as one approaches $\dot\gamma_c$ from above \cite{newexpts}, but in any case is usually much bigger than $\dot\gamma_c^{-1}$, so that a large total strain $\gamma_c$ needs to be applied before the thickening occurs. (For example, $\tau_{lat} \simeq 100$s in \cite{Berret}.) The latency period increases with the gap width of the shear cell \cite{newexpts} and in some cases, before it is over, there is clear spatial inhomogeneity \cite{Berret,pine}. 

Even after the onset of the gel, a true steady state is not necessarily reached; the stress may continue to adjust slowly for a much longer time $\tau_{slow}$ (hours) before achieving steady state. A similarly slow time scale controls the disappearance of structural memory after the shear has been terminated \cite{Berret,latency}; if shear is restarted within a shorter time, the latency time is eliminated or reduced. The latency time $\tau_{lat}$ can also be strongly sensitive to pretreatment of the sample: e.g., of samples sheared at $25\deg$C, a ten-fold increase of $\tau_{lat}$ was found in ones which had been held at a higher temperature for two hours beforehand \cite{Berret}. 

The existence of a slow relaxation time $\tau_{slow}$ is a widespread feature \cite{hoffmann}, recently confirmed for several new systems \cite{newexpts} such as gemini and fluorinated surfactants both with and without slight added salt. Importantly, recent work shows that much of the slow relaxation behavior observed in response to rheological disturbances can be reproduced by studying temperature changes (or even concentration changes \cite{concepts}) instead \cite{newexpts}. This confirms that some important timescales for rheology stem directly from the micellar kinetics in the quiescent state, as probed in $T$-jump (or sudden dilution). One of the effects of varying temperature is to alter $\bar L$ (the mean chain length) which shifts $\phi^*$ relative to the volume fraction $\phi$ of the sample. If rheological disturbances at relatively low shear rates can influence structure analagously, several characteristics of the shear thickening transition might be explained.

\section{A Ring-Driven Scenario}
Let us first assume that, near $\phi^*$, micellar rings form a percolating linked network in the quiescent state. It is not known whether or not this is truly the case, but the marginally overlapped structure predicted near the polymerization transition (at large enough $E$) ensures that the percolation transition is, at least, nearby. 

The system of linked rings would then form a solid gel were it not for the reaction kinetics of their linking and delinking. The delinking time $\tau_{link}$ can be much longer than the Zimm time of the rings (or any linear micelles that are present); fractions of a second are plausible. Such a long relaxation time would lead to a zero-shear viscosity enhancement; for a marginally connected ring network this is of order $\Delta\eta \sim n\tau_{link}$, where $n_l$ is the number of linked rings in the percolating network, per unit volume. For unlinked objects there is already, at $\phi^*$, roughly a factor two viscosity enhancement, arising from $\Delta\eta \sim n\tau_z$ where $n$ is the number of these objects. Therefore, existence of a linked network could go undetected in the zero shear viscosity so long as $n_l/n \le \tau_z/\tau_{link}$. If this amplitude is small (which, close to a percolation transition, it is) then the longer of the two relaxation times might also go undetected in, for example, transient birefringence and light scattering measurements \cite{tb}, but more careful examination might reveal a contribution of large, slow moving entities \cite{newexpts}. In principle the dramatic rise in the quiescent-state viscosity $\eta_0$ `above $\phi^*$' could arise from increasing connectivity of the linked network prior to the onset of the entangled condensate of long linear chains; this interpretation would place the ring percolation threshold slightly below $\phi^*$ as usually defined (by the minimum of collective diffusion or by a factor two increase in $\eta_0$).

If a linked ring network is indeed present in the quiescent state, then for shear rates above $\dot\gamma_c\simeq 1/\tau_{link}$, this would start to become strongly distorted and would soon dominate the stress, providing a basic mechanism for shear thickening, at shear rates low compared to the inverse Zimm time of the micelles. However, to explain its sharp onset around $\dot\gamma_c$, and the near-zero extinction angle of the shear-induced state, we also require a positive feedback mechanism between the strain of the linked network and its connectivity. Above $\dot\gamma_c$ this would explain the buildup, over the latency time $\tau_{link}$, of a highly aligned state. Given that the gel state is fully aligned (near zero extinction angle) the {\em shortest possible} `latency' time before reaching the steady state is the time taken to stretch a ring to full alignment. This requires a strain of order $N^{1/2}$, where $N\simeq \bar L/\ell$ is the mean number of persistence lengths in a micelle. For large $N$, this already gives a longish latency time: $\dot\gamma_c\tau_{lat}\sim N^{1/2}$. But this is a lower bound, which assumes perfect feedback so that in effect, for $\dot\gamma > \dot\gamma_c$, the delinking process becomes completely ineffective at relaxing ring orientations. In practice, any feedback mechanism will be gradual and the latency time will be longer. If the feedback does cause creation of a more highly linked (and then aligned) network, one can imagine this might propagate progressively through a Couette cell from regions of higher to lower shear rate giving a gap-dependent latency time \cite{newexpts} and transiently inhomogeneous birefringence \cite{Berret}.

What is the feedback mechanism? Let us argue that the distortion of the linked rings, when $\dot\gamma\tau_{link}\simeq 1$, causes changes in the statistics of their reaction kinetics. Suppose, for example, that the rings react purely by bond interchange. This means that a ring can fission into two smaller rings; any two rings can recombine to make a large one. 

The initial effect of straining a ring is to decrease the probability of it contacting itself (required for ring fission) relative to another ring (required for ring fusion); however this will be reversed at much higher strains where self-entanglement and finite extensibility requires the two limbs of the stretched ring to move towards each other. Around $\dot\gamma\tau_{link} \simeq 1$, where order-unity ring strains should arise, there is an enhanced tendency of fusion over fission, and the average ring size will increase. This can increase the connectivity of the linked network (positive feedback), causing an increased viscosity and a larger distortion per ring at the given $\dot\gamma$. The simplest model would then be an initial stress evolution, at fixed strain rate, of $\dot\sigma = \alpha\sigma$ where the feedback parameter changes sign at $\dot\gamma_c$ (e.g., $\alpha \sim \tau_{link}^{-1}(\dot\gamma-\dot\gamma_c)$). This gives a natural mechanism for large $\tau_{lat}$ near the transition. 

The two other main reaction schemes (end-interchange or reversible scission \cite{cc}) both require, as reagents and/or products, open linear chains as well as rings. (Delinking then proceeds via ring opening.) In each of these cases, the shear should again at first favour hetero-reaction over self-reaction \cite{witten}, but this now pushes the balance away from rings and toward chains. At first sight, this should destroy, not enhance, the linked network connectivity. However, near $\phi^*$, the opening of rings to make chains may itself be enough to explain a significant increase in viscosity, due to their increased end-to-end distance with increased overlap and mutual entanglement. This can increase the stress and promote stronger network alignment, possibly to the point where self-reaction to produce large rings is again favoured. Alternatively the new chains could shear band and/or create a nematic gel \cite{nemband}.

Without speculating further, we may conclude that, whenever percolating rings are present in the quiescent state with a delinking time $\tau_{link} \gg \tau_z$, then shear rates of order $\dot\gamma \sim 1/\tau_{link}$ are sufficient to disturb the kinetics of the system, with generally unpredictable consequences. These may  -- at least for some reaction schemes -- include runaway gelation to form an aligned network (or perhaps nematic) component. 

Accepting this, we may now relax our assumption that the rings already form a percolating network in the quiescent state. All that is needed is for flow rates to couple to the linking and delinking kinetics at some modest shear rate $\dot\gamma_c\tau \simeq 1$, with $\tau \gg \tau_z$. But for large $\tau_{link}$, to achieve this one needs only a partially percolating structure with some large fragments having $\tau_z \ll \tau \le \tau_{link}$. (For large $\xi$, the power law tail on the ring size distribution might be enough in itself.) In this case,
percolation of linked rings would only be completed as shear thickening proceeds. This is somewhat analagous to a {\em stress-induced} downward shift in the overlap threshold $\phi^*$ (more precisely, the percolation threshold) of the micellar rings. (See \cite{witten} for a related shear thickening mechanism in associating polymers.) Apart from its direct effects, causing strong network alignment, the effect of stress on structure could be quite similar to those caused by a decrease in temperature or an increase in concentration. 

\section{Routes to the steady state}
In steady state, a new balance between shearing and reactions must be struck. Depending on the kinetics, there are many possibilities for limiting the positive feedback cycle outlined above. For example, if reversible scission is the fastest reaction, shear-rates of order $\dot\gamma\simeq \tau_{link}^{-1}$ could open most of the rings; the resulting chains would become entangled (explaining a factor ten viscosity increase) but thereafter the effect of shear self-limiting, and the behavior of a normal viscoelastic micellar system recovered -- possibly including creation of a nematic phase  with near-zero extinction angle \cite{nemband}.

Another interpretation of the low extinction angle is that, for $\dot\gamma \ge \dot\gamma_c$, the feedback between strain and connectivity remains positive until full extension is approached. (A gaussian ring reaches this at a strain of order $N^{1/2}$ with extinction angle
$N^{-1/2}$.) At this point, the ongoing deformation of a  percolating linked network
(pre-existing or shear-induced) would force delinking. In principle this could be accomplished purely by bond interchange, but some flow-induced scission of rings to form open chains can probably be expected as well.

Any steady state would then comprise a mixture of rings and chains in various states of deformation, linkage, and alignment. An aligned ring network would dominate the birefringence and provide a very large first normal stress difference but (if the extinction angle is effectively zero) it would contribute little to the shear stress. The latter could be dominated by unlinked, though probably entangled, material comprising open chains and unlinked rings. Even if the medium remains spatially homogeneous throughout the process, such a balance may take some time to be reached. In the steady state, each ring in the aligned component eventually breaks, allowing relatively fast orientational relaxation (whether as a chain or a ring); the relaxed material sooner or later relinks into the network and enters a new deformation and breakage cycle. With such a mechanism, periodic or chaotic oscillations of the entire network (either transient \cite{Berret} or persistent \cite{rheochaos}) are not ruled out, but nor are they mandatory; the same remarks apply to spatial inhomogeneities \cite{pine,Berret}.

\section{Slow relaxations}
In $T$-jump measurements on open chains (no rings present), even if there are very fast bond- or end-interchange reactions, reversible scission alone controls the longest relaxation time for the chain size distribution \cite{jumprev}: only scission can change the total number of chains present in the system. In some cases, the scission process is too slow to be observed directly by $T$-jump \cite{candaujump}; a timescale (of minutes or hours) must then be inferred.
 
Interestingly, the same conservation law for interchange kinetics arises in a system of rings and chains. Bond interchange can create two rings out of one (or vice versa); end interchange can create one ring and one chain out of a chain only (or vice versa); and whichever of the three reactions is fastest will control $\tau_{link}$. However, only reversible scission allows the number of chains present to alter; it is therefore a {\em necessary} step in the relaxation process, whenever the balance between rings and chains has been perturbed. Such a perturbation can be induced by the onset of shearing, but it can also be induced by thermal cycling or any other preparative step which alters the number of open chains in the system. If the other reactions are fast, the effect of having a temporarily conserved (nonequilibrium) chain number is, without flow at least, simply to replace $E$ by a different value $\tilde E$ (which absorbs a Lagrange multiplier). This evolves slowly back towards $E$ on the timescale of scission.

Reversible scission is therefore a good candidate to explain the slow relaxation time $\tau_{slow}$. This timescale will only be well separated from $\tau_{link}$ if one of the other two reaction channels is much faster than scission;
this would allow the slow dynamics to be seen in some shear thickening micellar systems \cite{Berret,newexpts}, without requiring it in others. 
Crucially, this mechanism can explain a strong dependence of shear-thickening behaviour on thermal cycling and other aspects of sample history (over time scales $\tau_{slow}$) even in samples which have never previously undergone a shear-induced gelation. For example a thermal pretreatment at much higher temperature (reduced $E/kT$) will create more chains and leave fewer rings, in a less linked state; this could explain an increased latency time \cite{Berret}. (Put differently, one can expect to raise $\dot\gamma_c$, giving a larger $\tau_{lat}$ at any given $\dot\gamma$ \cite{newexpts}.) But unless by accident the number of chains created by the thermal cycling matches that required for the sheared steady state at the final temperature, there will still be a slow relaxation towards this state.

Pre-shearing a sample creates, in contrast, a population where the ring/chain balance is already close to the final one; so when shear at the same rate is re-started (after complete relaxation of stress, but long before the time scale $\tau_{slow}$ for equilibration of the chain number) the final steady state is recovered almost directly after the latency time $\tau_{lat}$ \cite{Berret}. But if this is the correct explanation of such data, a slow transient should be recovered if one preshears instead {\em at the wrong rate}. This could distinguish our picture from a nucleation mechanism in which any memory of previous shear resides in a few small droplets that can act as nucleation centres for the next cycle. It would be hard to see how the late stage growth of the new phase, from such nuclei, could be altered by the shear rate at which they had been created originally.

As already outlined, if rings at full extension undergo forced conversion into linear chains, then the final steady state will not be achieved until the flux of chains back into rings comes into balance. This may require a significant buildup in the number of open chains; if these enhance the shear stress, that may continue to rise until balance is finally restored on the time scale $\tau_{slow}$. (During this time, the birefringence should drop slightly; this might be worth checking experimentally.) However, if for some reason a slight excess of open chains is created early on, the slow transient should have the opposite sign. (We cannot explain how these two outcomes can result from identical sample preparation and flow history as reported by Berret et al \cite{Berret}.) On the other hand, if there is a vast excess of open chains (after a brutal heat-treatment) these will equilibrate among themselves with a reduced mean length (small $\tilde E$) so that the initial trend could again be a slowly rising shear stress, as the chains get longer. In fact, if the chain number is the only parameter controlling the transient, this rise should be followed by a maximum when relatively few excess chains are left, before the final plateau is again approached from above. This could also be worth checking by experiment.

\section{Role of added salt}
Within our picture, there are two pre-requisites for an abrupt shear thickening transition to be seen at modest shear rates; these are (a) the presence of rings near $\phi^*$, and (b) large values of $\tau_{link}$ ($\gg\tau_z$). A positive feedback between stress and ring linking is also needed; bond interchange offers the simplest mechanism for this, with other reaction schemes less predictable. Beyond this, to explain the slow transients sometimes seen, we also require reversible scission to be very slow (hours).

The presence of rings around $\phi^*$ requires large $E$. For a given chain structure the value of $\phi^*$ will presumably {\em exceed} the value $\phi_c$ for the underlying polymerization transition; to maximize the effect rings we should make $\phi^*$ as small, and $\phi_c$ as large, as possible.
Factors favouring large $E$ (small $\phi^*$) include raising the counterion lipophilicity and using gemini surfactants; increased micellar flexibility should raise $\phi_c$ \cite{catesring}. Such factors do seem broadly to correlate with low $\dot\gamma_c$ and a large viscosity increment \cite{newexpts}. 

In static terms, adding salt also raises $E$ (by reducing the electrostatic stabilization of end caps) and slightly decreases $\ell$. Why then, do moderate amounts of added salt generally {\em suppress} the shear-thickening transition? We argue that this is not through statics, but through the micellar kinetics. While destabilizing end-caps relative to the central cylinder, adding salt should also reduce the activation energy for all three reaction schemes, both through moderation of the the Coulomb barrier, and through relative stabilization of the inverted curvature required for the transition state; there is some preliminary evidence for this trend \cite{newexpts}. As soon as one of the reactions becomes fast enough that $\tau_{link}$ approaches $\tau_z$ (milliseconds or less) the interlinking of rings will become effectively invisible and our shear thickening scenario should fade away. Alternatively, one might propose that the strain/linking feedback mechanism can only occur when $\tau_{link}$ is governed by a specific reaction scheme, and that on adding salt, the relative rates are inverted so that another scheme takes over. 

\section{Conclusion}
This paper has outlined a speculative scenario for ring-driven shear thickening in micellar systems. One `prediction', which appears borne out by experiment, is that the details of the shear thickening behaviour can be quite complex and can vary strongly from one system to another. We would attribute this variation to (a) dependence on the quiescent ring/chain balance (factors: $E$, $\ell$, and sample history over the preceding interval $\tau_{slow}$); and (b) dependence on the relative rates of the various scission and interchange reactions (factors: as before, plus packing and Coulomb barriers in the activated states). The complexity of this basic picture could be compounded further if some sort of phase separation and/or nematic instability happens concurrently with the structural alterations we propose \cite{Berret,nemband}.

Such a radical scenario should not be accepted without significant further experiments to corroborate some of its predictions (and perhaps refine others; many details could be incorrect even if the basic idea is sound). This effort has partly been initiated already \cite{newexpts}, but a goal of this paper is to stimulate such an effort across a wider community, addressing a broader class of materials. Currently we are hopeful at least that {\em some elements} of our ring-driven shear thickening scenario may indeed apply to {\em some classes} of micellar materials. 

\acknowledgments
We thank P. Olmsted, M. Fuchs, R.M.L. Evans, C. Oeschlager, G. Waton and E. Buhler for illuminating discussions.

\end{document}